\documentclass[superscriptaddress,aps,journal=prl,twocolumn]{revtex4-2}
\usepackage{graphicx}
\usepackage{epstopdf}
\usepackage{bm,amsmath,amssymb} 
\usepackage{color, soul} 
\usepackage{dcolumn}   
\usepackage{multirow}

\begin{document}

\title{Two-dimensional ferroelectric metal for electrocatalysis}

\author{Changming Ke}
\affiliation{School of Science, Westlake University, Hangzhou, Zhejiang 310024, China}
\affiliation{Institute of Natural Sciences, Westlake Institute for Advanced Study, Hangzhou, Zhejiang 310024, China}
\affiliation{Key Laboratory for Quantum Materials of Zhejiang Province, Hangzhou Zhejiang 310024, China}
\author{Jiawei Huang}
\affiliation{School of Science, Westlake University, Hangzhou, Zhejiang 310024, China}
\affiliation{Zhejiang University, Hangzhou, Zhejiang 310058, China}
\author{Shi Liu}
\email{liushi@westlake.edu.cn}
\affiliation{School of Science, Westlake University, Hangzhou, Zhejiang 310024, China}
\affiliation{Institute of Natural Sciences, Westlake Institute for Advanced Study, Hangzhou, Zhejiang 310024, China}
\affiliation{Key Laboratory for Quantum Materials of Zhejiang Province, Hangzhou Zhejiang 310024, China}


\begin{abstract}{ 
The coexistence of metallicity and ferroelectricity has been an intriguing and controversial phenomenon as these two material properties are considered incompatible in bulk. We clarify the concept of ferroelectric metal by revisiting the original definitions for ferroelectric and metal. Two-dimensional (2D) ferroelectrics with out-of-plane polarization can be engineered via layer stacking to a genuine ferroelectric metal characterized by switchable polarization and non-zero density of states at the Fermi level. We demonstrate that 2D ferroelectric metals can serve as electrically-tunable, high-quality electrocatalysts. }
\end{abstract}
\maketitle
\newpage

The combination of seemingly incompatible material properties often leads to the discovery of novel emergent phenomena and unprecedented functional properties not available to conventional materials~\cite{Wu15p14501,Takabatake14p669,Qin17pe1601536}. The ``ferroelectric metal", first proposed by Anderson and Blount~\cite{Anderson65p217}, is such an example, and it refers to a material that is both ferroelectric and metallic, likely affording unique physical properties such as ferroelectric superconductivity~\cite{Rischau17p643} and magnetoelectric coupling~\cite{Edelstein95p2004}. In recent years, both experimental and theoretical studies have shown that atomic distortions that break the inversion symmetry can coexist with metallicity~\cite{Kolodiazhnyi10p147602, Wang12p247601,Shi13p1024,Fujioka15p13207,Puggioni15p087202,Kim16p68,Filippetti16p11211,Benedek16p4000,Cao18p1547,Fei18p336,Lu19p227601,Sharma19peaax5080}. Meanwhile, ferroelectric metal, together with other terms, e.g., ``noncentrosymmetric metal"~\cite{Puggioni14p3432}, ``polar metal"~\cite{Kim16p68}, ``metallic ferroelectrics"~\cite{Yimer20p174105}, and ``ferroelectric-like metal"~\cite{Fujioka15p13207}, were used somewhat interchangeably in literature to describe conductive materials without inversion symmetry, which has caused much confusion and debate~\cite{Hoch19p267}.

We first clarify the definition of ferroelectric metal by revisiting the original defintions for ``ferroelectric" and ``metal", respectively. The textbook definition for a metal is rather empirical: a substance with high electrical conductivity that decreases with increasing temperature~\cite{Ashcroft76}. Within the framework of single-particle band structure theory, the (semi)metal is a substance with nonzero density of states (DOS) at the Fermi level ($E_f$) at 0~K; this serves as a criteria for the intrinsic metallicity and is amenable to first-principles calculations. In practice, doping has been routinely used to tune the value of $E_f$ and to drive a band-gap insulator to a metallic state. For example, BaTiO$_{3-\delta}$ becomes metallic above a critical electron concentration of $n_c\approx 1\times 10^{20}$~cm$^{-3}$ whereas the cooperative atomic distortions retain until $n_c \gtrapprox 1.9\times 10^{21}$~cm$^{-3}$~\cite{Kolodiazhnyi10p147602}. In a rigid band picture, the metallicity of BaTiO$_{3-\delta}$ is due to the shift of $E_f$ and the valence band and the conduction band are still fully gaped in the Brillouin zone; this point will become relevant in defining the electric polarization in crystal with Born-von Karman periodic boundary conditions ~\cite{Ashcroft76} as discussed below. 

The definition of ferroelectricity was formulated based mostly on experimental observations. According to the standard reference book by Lines and Glass~\cite{Lines77}, a ferroelectric has two or more discrete orientational states, each associated with a nonzero electric polarization ($\mathcal{P}$) in the absence of an electric field ($\mathcal{E}$), and can be switched between these states via an electric field. It is interesting to note that the original definition of ferroelectricity did not rule out the coexistence of metallicity, at least not explicitly. Lines and Glass emphasized that the defining feature that distinguishes a ferroelectric from a pyroelectric is the {\em polarization reversibility} characterized by the $\mathcal{P}$--$\mathcal{E}$ hysteresis loop.
When measuring the hysteresis loop is difficult~\cite{Loidl08p191001}, the occurrence of a well-defined (displacive) phase transition to a nonpolar high-symmetry phase was also considered an indicator of ferroelectricity. In a similar spirit, the energy difference between the polar phase and the nonpolar reference phase computed with quantum mechanical methods was used to gauge the switchability~\cite{Garrity18p024115}. After revisiting the history, we arrive at the following definition based on microscopic materials descriptors: {\em a ferroelectric metal has non-zero density of states at the Fermi level and switchable two or more discrete polarization states, each being stable at null electric field.} We believe the confusions and debates regarding ferroelectric metal essentially boil down to two questions. (1) Can a metal have a well-defined electric polarization? (2) Is such electric polarization reversible?

\begin{figure}[t]
\centering
\includegraphics[scale=1]{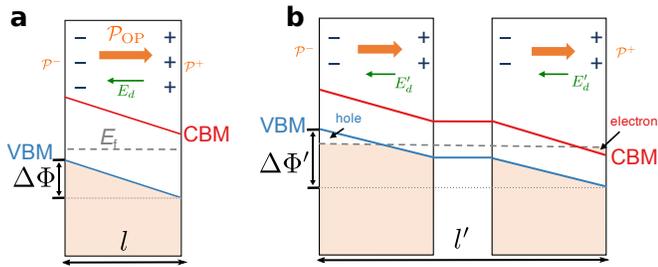}
 \caption{ Design principle for a two-dimensional ferroelectric metal. Band bending in ({\bf a}) monolayer and ({\bf b}) bilayer 2D ferroelectric with out-of-plane polarization ($\mathcal{P}_{\rm OP}$). The depolarizaiton field ($E_d$) due to incomplete screening creates a potential step ($\Delta \Phi$) such that the valence band minimum  (VBM) is located on the $\mathcal{P}^-$ surface and the conduction band maximum (CBM) is located on the $\mathcal{P}^+$ surface. Though $E_d' < E_d$ due to the stronger screening effect, the increased thickness ($l'$) of a bilayer in principle can lead to a larger $\Delta \Phi'\approx l'E_d'$ that drives the inversion of VBM and CBM. }
  \label{design}
 \end{figure}

For bulk materials, it is challenging to address these two questions convincingly via typical experimental techniques: the mobile carriers can strongly screen the external electric field, hindering the polarization reversal; a highly leaky sample is unlikely to result in a reliable $\mathcal{P}$--$\mathcal{E}$ loop; the ferroelectric-like hysteresis from piezoresponse force microscopy is not always reliable due to various non-ferroelectric origins such as electrochemical effects and charge accumulation~\cite{Kim16p102901}. LiOsO$_3$ was regarded as a prototypical ferroelectric metal because of a temperature-driven $R3c\rightarrow R\bar{3}c$ structural transition, though the spontaneous polarization was not observed~\cite{Shi13p1024}. Theoretically, the modern theory of polarization is based on Berry phase which is only well defined for a bundle of Bloch bands being an isolated subgroup~\cite{Smith93p1651, Smith94p5828,Resta93p133, Resta94p899}, thus limiting its application to insulators. It is possible to compute the Berry phase polarization (with some corrections accounting for the screening effect) for a semimetal with a nonzero direct band gap~\cite{Sharma19peaax5080,Watanabe20p104405}. For metallic systems with overlapping conduction and valence bands, however, it remains nontrivial to properly define $\mathcal{P}$ with the Berry phase approach~\cite{Filippetti16p11211}. Following that, the coupling term, $-\mathcal{E}\mathcal{P}$, becomes problematic, complicating the theoretical prediction of polarization reversibility. Therefore, we suggest it is more appropriate to categorize a metallic material in a polar space group as {\em polar metal} when the switchability remains unresolved. Here, {\em polar} means {\em polar space group}, not {\em polarization}. The name of ``noncentrosymmetric metal" is recommended to avoid confusion for metals lacking inversion symmetry but not in a polar space group. 

\begin{figure}[b]
\centering
\includegraphics[scale=1]{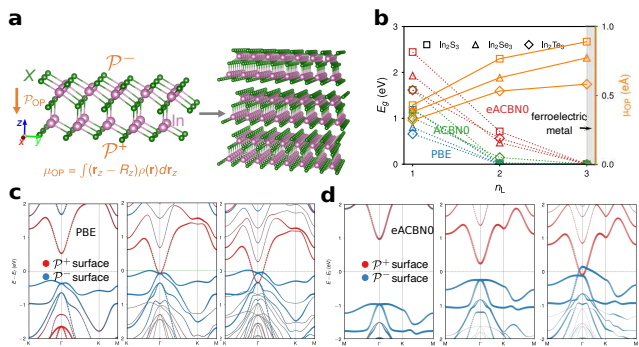}
 \caption{Electronic structures of multilayer $\alpha$-In$_2X_3$. ({\bf a}) Schematics of $\alpha$-In$_2X_3$ monolayer and stacked van der Waals structure. The out-of-plane electric dipole moment $\mu_{\rm OP}$ is obtained from the asymmetry of the total charge density ($\rho$) along $z$ with respect to the $z$-component of the geometry center $R_z$. ({\bf b}) Evolutions of PBE, ACBN0, and eACBN0 band gaps ($E_g$) and $\mu_{\rm OP}$ of a $2\times 2$ supercell as a function of layer number ($n_{\rm L}$). Projected band structures for 1L, 2L, and 3L $\alpha$-In$_2$Se$_3$ showing atomic orbital contributions from $\mathcal{P}^+$ and $\mathcal{P}^-$ surfaces obtained with ({\bf c}) PBE and ({\bf d}) eACBN0.}
  \label{bandstructure}
 \end{figure}

In this work, we propose the concept of ferroelectric metal can be defined rigorously for low-dimensional materials. For example, the two- or three-layer semi-metal WTe$_2$ with out-of-plane polarization ($\mathcal{P}_{\rm OP}$) and demonstrated switchability using gate electrodes is a true ferroelectric metal~\cite{Fei18p336}. Unlike the macroscopic polarization of a crystalline solid being well defined only for an insulator, the polarization of a bounded and charge-neural sample is a trivial quantity for both insulator and metal, determined by the asymmetry of the charge distribution, that is, electric dipole moment ($\mu$)~\cite{Resta21p050901}. Therefore, two-dimensional (2D) ferroelectric semiconductors with out-of-plane polarization ($\mathcal{P}_{\rm OP}$) serves as an ideal platform to realize genuine ferroelectric metal. Our design principle as illustrated in Fig.\,1 makes use of the potential step ($\Delta \Phi$) resulted from the built-in depolarization field ($E_d$) running against $\mathcal{P}_{\rm OP}$ to induce intrinsic metallicity. The valence band minimum (VBM) and the conduction band maximum (CBM) are located on the $\mathcal{P}^-$ and $\mathcal{P}^+$ surfaces, respectively. { A heuristic way to understand such band bending is that the system has the tendency to generate free carriers to compensate the surface bound charges. The VBM (CBM) moves toward the Fermi level so as to generate free holes (electrons) needed to compensate the $\mathcal{P}^-$ ($\mathcal{P}^+$) surface. Another way to comprehend the band diagram is that electrons close to the $\mathcal{P}^-$ surface naturally have higher energies because of the Coulomb repulsion. The depolarization field $E_d$ acts like the electric field within the depletion region of a $p$-$n$ junction (see Fig.~S7 in Supplemental Material).} By stacking 2D ferroelectrics, it is possible to further {increase $\Delta \Phi$ and} reduce the gap between VBM and CBMreduce the gap between VBM and CBM, eventually leading to a metallic state characterized by conductive holes on the $\mathcal{P}^-$ surface and electrons on the $\mathcal{P}^+$ surface. Because the whole system is finite along the out-of-plane direction, the value of $\mathcal{P}_{\rm OP}$, trivially related to $\mu_{\rm OP}$, can be computed from the asymmetry of the total charge density ({Fig.\,\ref{bandstructure}a}), free of the Berry phase issue discussed above. The reversibility of $\mathcal{P}_{\rm OP}$ can be evaluated by quantifying the field-induced changes in energy and atomic forces~\cite{Lu19p227601}.

We demonstrate the feasibility of the design principle using monolayer, bilayer, and trilayer (1L, 2L, and 3L) $\alpha$-In$_2X_3$ ($X$ = S, Se, Te)~\cite{Ding17p14956} as examples. The out-of-plane polarization in monolayer $\alpha$-In$_2X_3$ results from the displacement of central $X$ layer relative to the top and bottom In-$X$ layers (Fig.\,\ref{bandstructure}a). The presence of $\mathcal{P}_{\rm OP}$ in $\alpha$-In$_2$Se$_3$ with a thickness down to 3~nm were already confirmed in experiments~\cite{Cui18p1253,Xue18p4976,Xiao18p227601, Poh18p6340,Xue18p1803738}. 
We note that the polar surfaces of $\alpha$-In$_2X_3$ are dangling-bond-free and intrinsically stable requiring no screening mechanisms. This is drastically different from polar slabs of conventional FEs such as BaTiO$_3$ where the discontinuity of crystal lattice at surfaces results in unstable dangling bonds that must be compensated in some way (e.g., non-stoichiometric surface reconstruction)~\cite{Kalinin18p036502, Watanabe21p2155, Levchenko08p256101}. Our investigations of defective bilayer $\alpha$-In$_2$Se$_3$ confirm the robustness of $\mathcal{P}_{\rm OP}$ and metallicity against surface vacancies (see details in Supplemental Material, Sec. \uppercase\expandafter{\romannumeral2}).

\begin{figure}[b]
\centering
\includegraphics[scale=1]{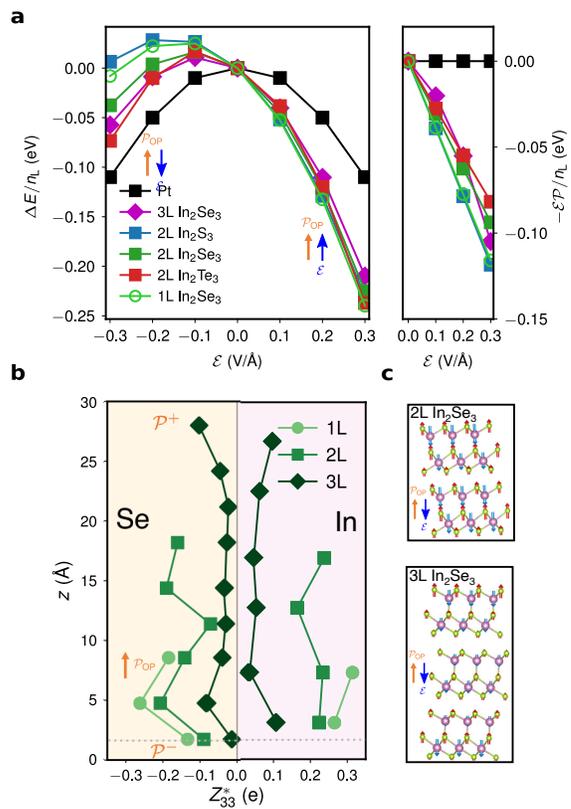}
 \caption{Polarization reversibility of metallic multilayer $\alpha$-In$_2X_3$. ({\bf a}) Effects of external electric fields applied along the out-of-plane direction on the energy change ($\Delta E$, left) and the electric field-polarization coupling strength ($-\mathcal{EP}$, right) per layer, compared with a Pt thin film. ({\bf b}) Born effective charges of Se and In atoms in 1L, 2L, and 3L $\alpha$-In$_2$Se$_3$. ({\bf c}) Atomic forces induced by an electric field of 0.3~V/\AA~applied against $\mathcal{P_{\rm OP}}$.}
  \label{Pswitch}
 \end{figure}
 
Density functional theory (DFT) calculations are performed with Vienna Ab initio Simulation Package (\texttt{VASP})~\cite{Kresse96p11169, Kresse96p15}. The exchange-correlation interaction is described by the Perdew–Burke–Ernzerhof (PBE) functional~\cite{Perdew96p3865} with D3 dispersion correction from Grimme~\cite{Grimme11p1456}. The projector augmented wave (PAW) method~\cite{Blochi94p17953} is used to describe the electron-ion interaction between the core ion and valence electrons. And we use an energy cutoff of 380~eV and a $7\times 7\times 1$ Monkhorst-Pack $k$-point mesh for structural optimizations with an ionic energy convergence threshold of $10^{-5}$~eV. The band structures are calculated using a denser $k$-point mesh and an energy cutoff of 500~eV. Spin polarization and dipole correction are considered in all calculations. The vacuum region is thicker than 10~\AA. To model external electric field, a planar dipole layer is introduced in the middle of the vacuum region in DFT calculations~\cite{Neugebauer92p16067}. The electrical conductivity and carrier mobility are estimated using the semiclassical Boltzmann transport theory within the constant relaxation time approximation, as implemented in the \texttt{BoltzTrap} code~\cite{Madsen06p67}, and a $20\times20\times 1$ $k$-point mesh is employed for the corresponding electronic structure calculations. Additional details regarding adsorption energy calculations are reported in Supplemental Material, Sec. \uppercase\expandafter{\romannumeral3}.

The band gap ($E_g$) reduction and closure driven by the increasing layer number ($n_{\rm L}$) is clearly visible in Fig.\,\ref{bandstructure}b. At the PBE level, all studied multilayers ($n_{\rm L}\geq 2$) are metallic. By the meanwhile, $\mu_{\rm OP}$ increases with $n_{\rm L}$ though the dipole per layer ($\mu/n_{\rm L}$) becomes smaller, suggesting a stronger screening effect.

It is well known that semi-local density functionals like PBE will underestimate $E_g$. Hybrid functionals with some percentage of exact exchange assuming fixed dielectric screening (e.g., HSE06) are also inappropriate for predicting band gaps of low-dimensional material systems since the rapid variation of screened Coulomb interactions is not captured~\cite{Jain11p216806}. To address this issue, we investigate the $n_L$-dependence of $E_g$ using newly developed pseudohybrid Hubbard density functional, Agapito-Cuetarolo-Buongiorno-Nardelli (ACBN0)~\cite{Agapito15p011006}, and the extended version (eACBN0)~\cite{Lee20p043410,Tancogne-Dejean20p155117}. The ACBN0 functional can be regarded as a DFT+$U$ method with the on-site Coulomb repulsion $U$ value determined self-consistently. The eACBN0 is an extended DFT+$U$+$V$ method with $V$ representing inter-site Coulomb repulsion  between the neighbouring Hubbard sites. The self-consistent calculations of $U$ and $V$ in eACBN0 are performed on each atom, thus capturing the local variations of screening of Coulomb interactions and leading to better descriptions of the electronic structures of low-dimensional materials than HSE06 and GW~\cite{Lee20p043410}. Our previous benchmark studies have shown that for monolayer and bilayer $\alpha$-In$_2$Se$_3$, eACBN0 predicts a larger band gap than ACBN0 and HSE06~\cite{Huang20p165157}. As shown in Fig.\,\ref{bandstructure}b, both 
ACBN0 and eACBN0 predict a decrease of $E_g$ with increasing $n_{\rm L}$ and a metallic trilayer.

The band structures of 1L, 2L, and 3L $\alpha$-In$_2$Se$_3$ with the atomic orbital contributions from $\mathcal{P}^+$ and  $\mathcal{P}^-$ surfaces obtained with PBE and eACBN0 are reported in Fig\,\ref{bandstructure}c and d, respectively. Consistent with the design principle, the VBM is dominated by states of $\mathcal{P}^-$ surface while the CBM is mainly comprised of states of $\mathcal{P}^+$ surface, and an increasing $n_{\rm L}$ reduces the energy difference between band edges,  eventually leading to a strong overlap between valence and conduction bands and a metallic state. Additional calculations also confirm the persistent metallicity in a realistic setup (e.g, in the presence of a polarizable medium represented by water, see Fig.\,S4 in Supplemental Material). Given the similar trend revealed by different DFT methods, we explore the main features of 2D ferroelectric metal and related functional properties (e.g., electrocatalysis) with PBE for its low computational cost.

We then investigate the polarization reversibility of multilayer $\alpha$-In$_2X_3$ by computing the energy change [$\Delta E = E(\mathcal{E})-E(\mathcal{E}=0)$] due to the application of an electric field along the out-of-plane direction. The atoms were fully relaxed in the presence of $\mathcal{E}$. We use a thin Pt slab consisted of three atomic layers ($\approx$ 4.9~\AA) as a reference. As shown in Fig.\,\ref{Pswitch}a, the $\Delta E$--$\mathcal{E}$ plot is symmetric for the Pt slab, indicating there is no preference of field direction. In comparison, the energy profiles for all three bilayers are asymmetric, and the structures with parallel $\mathcal{P}_{\rm OP}$ and $\mathcal{E}$ have lower energies. Similar to bulk ferroelectrics, the bilayers, albeit being metallic, exhibit a decent $\mathcal{P}$--$\mathcal{E}$ coupling strength (close to that in semiconducing monolayer In$_2$Se$_3$), which can serve as the driving force for polarization switching. 

For insulators, the Born effective charge (BEC) tensor ($Z^*_{\kappa, \beta \alpha}$) describes the linear relationship between the polarization per unit cell along the direction $\beta$ and the displacement ($u$) along the direction $\alpha$ of atoms in the sublattice $\kappa$; it is also the linear proportionality coefficient relating the atomic force $\mathcal{F}$ and the macroscopic electric field, $Z^*_{\kappa, \beta \alpha} = \Omega_0\partial{\mathcal{P}_{\beta}}/\partial{u_{\kappa\alpha}} = \partial{\mathcal{F_{\kappa\alpha}}}/\partial\mathcal{E}_{\beta}$, where $\Omega_0$ is the unit cell volume~\cite{Gonze97p10355}. Therefore, for 2D metallic systems explored here, we can define a generalized BEC $Z^*_{33} = \partial{\mathcal{F}_3}/\partial\mathcal{E}_{3}$ along the out-of-plane direction, which inversely scales with the degree of electric field screening of mobile carriers.  We compare the magnitudes of $Z^*_{33}$ of In and Se atoms in 1L, 2L, and 3L $\alpha$-In$_2$Se$_3$ in Fig.\,\ref{Pswitch}b.  Different from anonymously large BECs in bulk ferroelectrics such as BaTiO$_3$~\cite{Zhong94p3618}, the values of $Z^*_{33}$ in the monolayer are rather small, $\approx$$-0.20$~$e$ for Se and $\approx$$0.3$~$e$ for In. This explains the persistent $\mathcal{P}_{\rm OP}$ in 2D as small effective charges hint at both unstable transverse optic mode and longitudinal optic model in the corresponding high-symmetry nonpolar phase,  referred to as ``hyperferroelectricity"~\cite{Garrity14p127601}. Interestingly, the magnitudes of $Z^*_{33}$ in the metallic bilayer are only 25\% smaller than those in the semiconducting monolayer on average, demonstrating a relatively weak screening of conducting carriers and a feasible polarization reversal. The BECs in trilayer $\alpha$-In$_2$Se$_3$ still maintain $\approx$30\% of monolayer values, $\approx$$-0.04$~$e$ for Se and $\approx$$0.07$~$e$ for In. It is noted that even for atoms in the middle layer, the charges are not zero, different from previous studies on polar metal LiOsO$_3$ where the internal atoms in a four-unit cell-thick thin film have nearly zero BECs ($<$0.01~$e$)~\cite{Lu19p227601}. The computed field-induced forces (Fig.\,\ref{Pswitch}c) in 2L and 3L $\alpha$-In$_2$Se$_3$ show nearly all atoms are affected by the applied field. We also note that Se atoms on the $\mathcal{P}^+$ and $\mathcal{P}^-$ surfaces get screened differently with the latter almost fully screened, likely due to the subtle differences in effective mass and mobility of conducting electrons and holes.  

 \begin{figure}[t]
\centering
\includegraphics[scale=1]{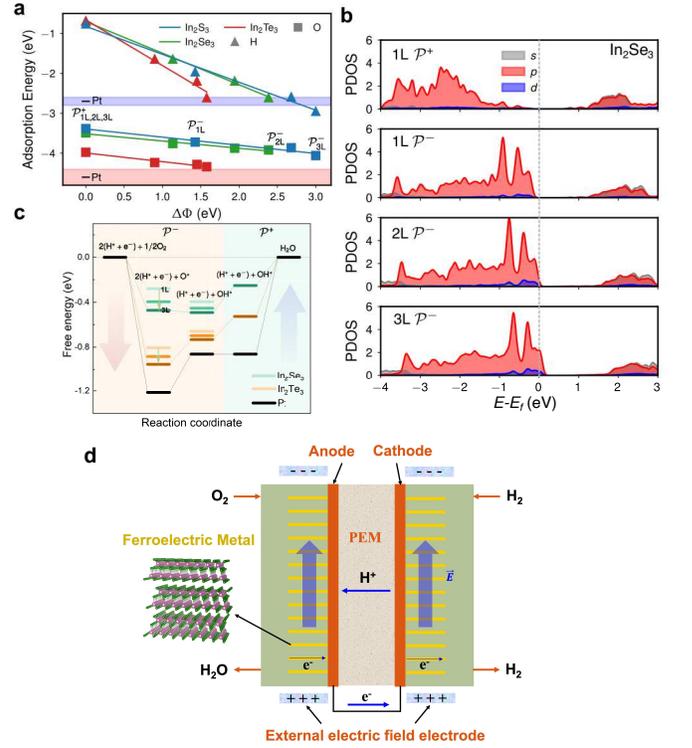}
 \caption{Electrocatalysis of 2D ferroelectric metal. ({\bf a}) Adsorption energies of H and O on the surfaces of multilayer $\alpha$-In$_2X_3$ uniaxially strained by 2\% as a function of potential step with a surface coverage of 12.5\%. The blue (red) shaded area shows the H (O) adsorption energy on Pt with a $\pm$0.1 ($\pm$0.2) eV energy window, indicating high activity region for hydrogen evolution (oxygen reduction) reaction \cite{Norskov05pJ23,Norskov04p17886}. ({\bf b}) Partial density of states of surface atoms in layered $\alpha$-In$_2$Se$_3$. ({\bf c}) Oxygen reduction reaction at the equilibrium potential 1.23 V enhanced by switchable electrocatalysis. Changing the reacting surface from $\mathcal{P}^-$ to  $\mathcal{P}^+$ will facilitate (H$^+$+e$^-$)+OH$^*$$\rightarrow$H$_2$O. {({\bf d}) Schematic diagram of hydrogen PEM fuel cell incorporating 2D ferroelectric metals for switchable electrocatalysis.}}
  \label{ads}
 \end{figure}
 
We have shown that the multilayer structure consisted of 2D ferroelectrics with $\mathcal{P}_{\rm OP}$ can be made into a genuine ferroelectric metal characterized by switchable polarization and nonzero density of states at $E_f$. We now discuss a potential application of ferroelectric metal overlooked in literature, that is, switchable electrocatalysis. The concept of ferroelectric catalysis is well established based on the fact that the surface structures and catalytic properties depend sensitively on the bulk polarization~\cite{Kolpak08p036102,Kakekhani16p19676,Efe21p024702,Wan21p7096}. {By applying a cyclic electric field to switch between one surface state with strong adsorption potential and another surface state with strong desorption potential, it is possible to overcome some of the fundamental limitations on the catalytic activity imposed by the Sabatier principle~\cite{Medford15, Kakekhani15p4537,Kakekhani16p302}. Photocatalysis accelerated by switchable chemistry of ferroelectrics has been reported in a few reactions such as direct decomposition of NO into N$_2$ and O$_2$ on CrO$_2$/PbTiO$_3$~\cite{Kakekhani15p4537} and water splitting on PbTiO$_3$~\cite{Kakekhani16p5235}.} However, the issues of poor charge transport and low conductivity of wide-band-gap ferroelectrics make them unsuitable for the electrocatalytic reaction to occur. Our calculations reveal that layered $\alpha$-In$_2$Se$_3$ have improved electrical conductivity (see Fig.\,S6 in Supplemental Material) thus potentially enabling electrocatalysis with tunable carrier type.

Moreover, we find that the carrier mobility estimated with semiclassical Boltzmann transport equation increases with $n_{\rm L}$, 61, 432, 695 cm$^2$/V/s in 1L, 2L, and 3L, respectively, much higher than typical metals such as Al and Cu (see Table S4 in Supplemental Material). The highly mobile carriers are beneficial for enhancing the kinetics of electrocatalytic reactions. 

Detailed mechanistic studies of specific electrocatalytic reactions are beyond the scope of current work. Here, we offer a proof-of-principle study on the adsorption energies of H and O atoms on stacked $\alpha$-In$_2X_3$. As shown in Fig.\,\ref{ads}a, both H and O atoms have much stronger binding energies on the $\mathcal{P}^-$ surface (where the VBM resides), and the adsorption energy increases linearly with $n_{\rm L}$ and the corresponding potential step $\Delta \Phi$ across the multilayer structure (the potential at $\mathcal{P}^+$ surface is set as zero). It is noted that these transition-metal-free 2D materials have adsorption energies comparable with precious metals such as Pt (Fig.\,\ref{ads}a), which may broaden the materials design space for 2D electrolysis.  We find that the $n_{\rm L}$-dependent adsorption energy can not be explained with a simple electrostatic model considering only surface charges (see discussions in Supplemental Material, Sec. \uppercase\expandafter{\romannumeral7} ). The trend can be understood by generalizing the famous $d$-band theory~\cite{ Hammer95p238} which states that the higher the $d$-band center, the stronger the interaction between the adsorbate and the metal surface. The partial density of states of surface layers (Fig.\,\ref{ads}b) reveal that the states near $E_f$ are mostly of $p$-orbital character, and an increasing $n_{\rm L}$ and $\Delta \Phi$ effectively pushes the $p$-band center higher and results in stronger bonding. 
Because $\Delta \Phi$ can be tuned continuously by varying the magnitude of external $\mathcal{E}$, the multilayer-based ferroelectric metal allows optimal design of adsorption energies by engineering the band center. For example, the oxygen reduction reaction (ORR) is one of the most important reactions in energy conversion systems such as fuel cells. Noble metal Pt, showing a high ORR reactivity with a moderate oxygen adsorption energy, is the most commonly used electrode material. As illustrated in Fig.\,\ref{ads}c, the reactivity of $\mathcal{P}^-$ surface of layered $\alpha$-In$_2X_3$ is approaching Pt with increasing $n_{\rm L}$. Furthermore, by switching the direction of the polarization, the final step, (H$^+$+e$^-$)+OH$^*$ $\rightarrow$ H$_2$O, that favors a low adsorption energy can be facilitated \cite{Michaelides03p3704}. {Finally, we suggest a structural design of hydrogen polymer electrolyte memebrane (PEM) fuel cell
incorporating ferroelectric metals, the performance of which can be readily tuned by applying an external electric field (Fig. 4d)}.

In summary, we demonstrate the concept of ferroelectric metal in layered van der Waals structure consisted of 2D ferroelectrics with out-of-plane polarization. The ferroelectricity and metallicity can coexist in low-dimensional materials with switchable polarization. The screening effect of mobile carriers is quantified with generalized Born effective charges computed from the field-induced atomic forces. In model ferroelectric metal represented by multilayer $\alpha$-In$_2X_3$, the mobile carriers can not fully screen the external electric field. The simultaneous presence of tunable polarization, mobile carriers, and electrical conductivity in two-dimensional ferroelectric metal offers a new platform to design transition-metal-free electrocatalysts with optimal adsorption energies.\\

\begin{acknowledgments}
C.K., J.H., S.L. acknowledge the supports from Westlake Education Foundation, Westlake Multidisciplinary Research Initiative Center, and National Natural Science Foundation of China (52002335). The computational resource is provided by Westlake HPC Center.
\end{acknowledgments}



\bibliography{SL}
\end{document}